\documentclass{PoS}

\usepackage{amsmath}
\usepackage{amssymb}
\usepackage{epsfig}
\usepackage{etex, xy}
\xyoption{all}

\newcommand{\ds}{\displaystyle}
\newcommand{\sign}[1]{\textnormal{sign}\hspace{-0.1em}\left(#1\right)}
\newcommand{\abs}[1]{\left|{#1}\right|}

\definecolor{blaugrau}{rgb}{0.796887, 0.789075, 0.871107}

\newcounter{mmacnt}
\def\restartmma{\setcounter{mmacnt}{0}}
\restartmma \catcode`|=\active
\def|#1|{\mathrm{#1}}
\catcode`|=12
\newenvironment{mma}{
 \par
 \catcode`|=\active
 \parskip=0pt\parindent=0pt 
 \small
 \def\In##1\\{%
   \def\linebreak{\hfill\break\null\qquad}%
   \refstepcounter{mmacnt}
   \hangindent=2.5em\hangafter=0
   \leavevmode
   \llap{\tiny\sffamily In[\arabic{mmacnt}]:=\kern.5em}%
   \mathversion{bold}\scriptsize$\tt\bf\displaystyle##1$\normalsize
   \mathversion{normal}\par
 }%
 \def\Print##1\\{%
   \def\linebreak{\hfill\break}%
   \hangindent=2.5em\hangafter=0
   \leavevmode\scriptsize ##1\par}%
 \def\Out##1\\{%
   \vspace*{-0.2cm}\def\linebreak{$\hfill\break\null\hfill$}%
   \kern\abovedisplayskip\par
   \hangindent=2.5em\hangafter=0
   \leavevmode
   \llap{\tiny\sffamily Out[\arabic{mmacnt}]=\kern.5em}
   \scriptsize$\displaystyle\tt##1$\normalsize\hfill\null\par
   \kern\belowdisplayskip\vspace*{-0.3cm}
 }%
 \def\Warning##1##2\\{%
   \def\linebreak{\hfill\break}%
   \hangindent=2.5em\hangafter=0
   \leavevmode
   {\scriptsize##1 : ##2}\par}%
}{%
 \par\smallskip
}

\newcommand{\LoadP}[1]{\fcolorbox{black}{blaugrau}{
\begin{minipage}[t]{10cm}
\footnotesize #1
\end{minipage}}}

\newcommand{\myIn}[1]{{\sffamily In[#1]}}
\newcommand{\myOut}[1]{{\sffamily Out[#1]}}

\def\MLabel#1{{\refstepcounter{mmacnt}\label{#1}}\addtocounter{mmacnt}{-1}}

\let\set\mathbb
\newcommand{\ep}{\varepsilon}

\newcommand{\NN}{\mathbb{N}}
\newcommand{\ZZ}{\mathbb{Z}}

\newcommand{\RR}{\mathbb{R}}
\newcommand{\KK}{\mathbb{K}}

\newcommand{\IIntegral}{I}

\title{
{\footnotesize\textnormal{DESY 16-003, DO-TH 16/01}}\\
A toolbox to solve coupled systems of differential and difference equations\thanks{This work was 
supported in part by
the Austrian Science Fund (FWF) grant SFB F50 (F5009-N15) and
the European Commission through
contract PITN-GA-2012-316704
({HIGGSTOOLS}).}}

\ShortTitle{A toolbox to solve coupled systems of differential and difference equations}

\author{Jakob Ablinger and \speaker{Carsten Schneider}
\\
        Research Institute for Symbolic Computation (RISC)\\
        Johannes Kepler University, Altenbergerstra\ss{}e 69, A-4040 Linz,
Austria\\
        E-mail: \email{Jakob Ablinger@risc.jku.at,Carsten.Schneider@risc.jku.at}}

\author{Johannes Bl\"umlein and Abilio de Freitas\\
        Deutsches Elektronen--Synchrotron, DESY,\\
Platanenallee 6, D--15738 Zeuthen, Germany\\
        E-mail: \email{johannes.bluemlein@desy.de,abilio.de.freitas@desy.de}}

\abstract{We present algorithms to solve coupled systems of linear differential equations, arising in the 
calculation of massive Feynman diagrams with local operator insertions at 3-loop order, which do {\it not} request special choices 
of bases. Here we 
assume that the desired solution has a power series representation and we seek for the coefficients in closed 
form. In particular, if the coefficients depend on a small parameter $\ep$ (the dimensional parameter), 
we assume that the coefficients themselves can be expanded in formal Laurent series w.r.t.\ $\ep$ and we try 
to compute the first terms in closed form. More precisely, we have a decision algorithm which solves the 
following problem: if the terms can be represented by an indefinite nested hypergeometric sum expression 
(covering as special cases the harmonic sums, cyclotomic sums, generalized harmonic sums or nested 
binomial sums), then we can calculate them. If the algorithm fails, we obtain a proof that the terms cannot 
be represented by the class of indefinite nested hypergeometric sum expressions. Internally, this problem 
is reduced by holonomic closure properties to solving a coupled system of linear difference equations. 
The underlying method in this setting relies on decoupling algorithms, difference ring algorithms and 
recurrence solving. We demonstrate by a concrete example how this algorithm can be applied with the new 
Mathematica package \texttt{SolveCoupledSystem} which is based on the packages \texttt{Sigma}, 
\texttt{HarmonicSums} and \texttt{OreSys}. In all applications the representation in $x$-space is obtained as an 
iterated integral representation over general alphabets, generalizing Poincar\'{e} iterated 
integrals.}

\FullConference{12th International Symposium on Radiative Corrections (Radcor 2015) and LoopFest XIV (Radiative Corrections for the LHC and Future Colliders)\\
15-19 June, 2015\\
UCLA Department of Physics \& Astronomy Los Angeles, USA
}

\begin{document}

\section{Introduction}

A massive three loop Feynman diagram $I(N)$ with a local operator insertion can be written in terms of multiple integrals or multiple sums, which depend 
on a discrete variable $N \in \mathbb{N}$ and the dimensional parameter $\varepsilon = D - 4$, 
where $D \in \mathbb{R}$ denotes the space-time dimension, see e.g. see~\cite{Blumlein:2010zv,Weinzierl:13}. Then one is interested in the first coefficients of its formal Laurent series  
w.r.t.\ $\ep$ (in short $\ep$-expansion)
\begin{equation}\label{Equ:IExpansion}
I(N)=\ep^{o}I_{o}(N)+\ep^{o+1}I_{o+1}(N)+\ep^{o+2}I_{o+2}(N)+\dots
\end{equation}
with order $o\in\ZZ$.
In this article we present tools to decide algorithmically if the coefficients $I_i(N)$ up to a certain order can be written in terms of indefinite nested hypergeometric sums (in short, nested hypergeometric sums) which can be defined as follows. Let $f(N)$ be an expression that evaluates at non-negative 
integers (from a certain point on) to elements of a field $\set K$ containing the rational numbers $\set Q$. Then $f(N)$ 
is called a nested hypergeometric sum expression w.r.t.\ $N$ if it is composed by elements from the rational function field $\set K(N)$, the three operations 
($+,-,\cdot$), hypergeometric expressions of the form $\prod_{k=l}^Nh(k)$ with $l\in\set N$ and $h(k)$ being a rational function in $k$ and being free of $N$, and sums of the form $\sum_{k=l}^Nh(k)$ with $l\in\set N$ and with 
$h(k)$ being a nested hypergeometric sum expression w.r.t.\ $k$ and being free of $N$. This class of special functions covers as special cases harmonic sums~\cite{BlumVerm},
\begin{eqnarray}\label{Equ:HarmonicSumsIntro}
	S_{a_1,\ldots ,a_k}(N)= \sum_{N\geq i_1 \geq i_2 \geq \cdots \geq i_k \geq 1} \frac{\sign{a_1}^{i_1}}{i_1^{\abs {a_1}}}\cdots
	\frac{\sign{a_k}^{i_k}}{i_k^{\abs {a_k}}}	
\end{eqnarray}
for non-negative integers $N$ and nonzero integers $a_i$ $(1 \leq
i \leq k)$ and more generally, generalized harmonic sums~\cite{Moch:2001zr,Ablinger:2013cf}, cyclotomic 
harmonic sums~\cite{Ablinger:2011te} or nested binomial 
sums~\cite{BinSums,Ablinger:2014bra}.\footnote{For surveys on these quantities see e.g. \cite{REF1}.}

In order to calculate the coefficients in~\eqref{Equ:IExpansion}, the summation package \texttt{Sigma} enhanced by the package \texttt{EvaluateMultiSums}~\cite{Summation,CASummation}, the integration package \texttt{MultiIntegrate}~\cite{Integration}, and the package \texttt{HarmonicSums}~\cite{HarmonicSums} have been applied successfully in many applications. However, in the course of recent calculations, these tools turned out to be not
sufficient and we extended them significantly by using the integration by parts 
(IBP) identities~\cite{IBP}.
Namely, encoding $I(N)$ by a generating function (formal power series) 
$$\hat{I}(x)=\sum_{N=0}^{\infty}I(N)\,x^N,$$ 
we can activate the powerful {\tt C++} program {\tt Reduze~2}~\cite{Reduze2} based on Laporta's algorithm \cite{Laporta:2001dd}
to reduce $\hat{I}(x)$ to a linear combination of master integrals. 
In many calculations, see~e.g.,~\cite{Ablinger:2014uka,DiffCalculations,VLadders} these remaining integrals are now suitable for symbolic summation and integration. However, some of the master integrals are rather hard too handle or are not in the proper form for symbolic summation and integration. Here we utilize the fact that  {\tt Reduze~2} can produce recursively defined coupled systems of linear differential equations in terms of these master integrals. Following the tactics in~\cite{DEQ} the main task is to extract the required information from these coupled systems and to reassemble information of the coefficients $I_i(N)$ in~\eqref{Equ:IExpansion} for further processing. In this article we are interested in computing the $I_i(N)$ in closed form. If a first-order coupled system has a specific form, one could use, e.g., the methodology described in \cite{Henn:2013pwa}. In the following we introduce a very general and efficient approach relying on decoupling algorithms~\cite{UNCOUPL,Zuercher:94,OreSys}, recurrence solvers~\cite{dAlembert,ParticularSol} and difference ring algorithms~\cite{DRTheory,Schneider:10b,DRTheoryMain}: we obtain a complete algorithm that extracts the first coefficients of the Laurent series and computes simultaneously the representation of the coefficients in terms of nested hypergeometric sum expressions, whenever this is possible. The first ideas of this new algorithm have been introduced in~\cite{NewUncouplingMethod} and the main features are worked out in~\cite{VLadders} by concrete examples coming from massive $3$-loop ladder and $V$-diagrams. In the following we will complement these achievements by precise input-output specifications and further details of the algorithms. Moreover, we will illustrate how the differential equation algorithm can be executed within the new package \texttt{SolveCoupledSystem} that relies on the packages \text{Sigma} and \texttt{OreSys}. In addition, the package \texttt{HarmonicSums} is used to gain significant speed-ups.

\medskip

We will use the following notations.
Let $\set K$ be a computable field containing the rational numbers as sub-field (e.g., $\set K=\set Q$). In the following $\KK[N]$ (or $\KK[\ep,N]$) denotes the ring of polynomials in the variable $N$ (or in the variables $\ep$ and $N$). Moreover, $\set K(N)$ (or~$\set K(\ep,N)$) denotes the field of rational functions in the variable $N$ (or in the variables $\ep$ and $N$).
We denote by $\KK((\ep))$ the field of formal Laurent series, i.e., elements are of the form $\sum_{k=o}^{\infty} f_k\ep^k$ with $f_k\in\KK$ and $o\in\ZZ$. 
Furthermore, we denote by $\KK((\ep))[[x]]$ the ring of power series whose elements are of the form $\sum_{i=0}^{\infty}f_i\,x^i$ with $f_i\in\KK((\ep))$.
Furthermore, $\KK^{\NN}$ (or $\KK((\ep))^{\NN}$) denotes the ring of sequences with entries form $\KK$ (or from $\KK((\ep))$).

\section{The algorithmic machinery}

First, we will address the problem how one can decide algorithmically, if a sequence can be calculated by a nested hypergeometric sum expression provided that the sequence is described by a linear recurrence (linear difference equation) in terms of nested hypergeometric sum expressions (see Theorem~1). Given this technology, we can extract the first coefficients of a Laurent series expansion in terms of nested hypergeometric sum expressions provided that the Laurent series is a solution of a recurrence of certain kind (see Theorem~2). Using uncoupling algorithms, this result can be generalized further to coupled systems (see Theorem~3). Finally, we can carry over this result to coupled systems of linear differential equations (see Theorem~4).

\subsection{Finding nested hypergeometric solutions of linear recurrences}\label{Subsec:RecSolver}

The main engine relies on the following algorithmic result~\cite{CASummation,DRTheoryMain,dAlembert,ParticularSol}.

\medskip

\noindent\textbf{Theorem 1.}
Suppose that a sequence $\langle I(N)\rangle_{N\geq0}\in\KK^{\NN}$
is a solution of the difference equation 
\begin{equation}\label{Equ:SimpleRec}
a_0(N)I(N)+a_1(N)I(N+1)+\dots+a_d(N)I(N+d)=r(N)
\end{equation}
with $N\geq0$ for given rational functions $a_0(N),\dots,a_d(N)\in\KK(N)$, not all zero, and a nested hypergeometric sum expression $r(N)$.
Then one can determine an $m\in\NN$ with the following property.\\
If one is given the values $I(N)$ for all $0\leq N\leq m$, then one can decide algorithmically if there exists a nested hypergeometric sum expression that calculates the values $I(N)$ for all $N\geq0$ 
(or at least from a certain point on).

\smallskip

\noindent\textit{Proof.} For this result we refer to Section~4.3 in~\cite{CASummation}
and Section~2.4~\cite{DRTheoryMain}. It has been implemented within \texttt{Sigma} as follows. The recurrence operator is factorized as much as possible into linear factors. Then each linear factor leads to one additional linearly independent solution of the homogeneous version of the recurrence by introducing one extra indefinite summation quantifier and introducing one hypergeometric expression~\cite{dAlembert}. In this way one finds a basis of all solutions of the homogeneous recurrence that can be expressed in terms of nested hypergeometric sum expressions. If the recurrence factorizes completely, the particular solution can be obtained straightforwardly. However, if the recurrence does not fully factorize, one has to activate algorithms from~\cite{ParticularSol} in order to calculate a particular solution in terms of nested hypergeometric sum expressions or to prove that such a representation is not possible. If there is not such a solution, $I(N)$ cannot be represented by a nested hypergeometric sum expression.\\ 
In the process of this calculation one can determine a $\mu\in\NN$ such that the solutions in terms of nested hypergeometric sum expressions can be evaluated and are a solution of the recurrence. Now compute the finite set $R$ of all non-negative integer roots of $a_d(N)\in\KK[N]$. If $R=\{\}$, set $\mu'=\mu$, else set $\mu':=\max(1+\max(R),\mu)$. Thus for all $\lambda\in\NN$ with $\lambda\geq\mu'$ we have that $a_d(\lambda)\neq0$.
Then using $d$ initial values, namely $I(\mu'),\dots,I(\mu'+d-1)$, one can check if the found solutions 
can be combined to an expression which produces the same initial values. If this is possible, this 
expression agrees with $I(N)$ for all $N\geq \mu'$: since the leading coefficient 
of~\eqref{Equ:SimpleRec} does not evaluate to zero, there is exactly one sequence which has these $d$ 
initial values and which is a solution of the recurrence. Note that this construction is always successful if one computes $d$ linearly independent solutions of the input recurrence (the particular solution is then just a by-product). Otherwise, if this construction fails, it follows that $I(N)$ cannot be expressed by a  nested hypergeometric sum expression (i.e., that at least one solution of the homogeneous recurrence is of different nature). Summarizing, if we can compute the first  $m:=\mu'+d-1$ initial values of $I(N)$, we can execute the decision procedure described above.

\smallskip

\noindent\textit{Example.} Consider the sequence $I(N)$ that is determined by the linear recurrence
\begin{mma}\MLabel{MMA:SimpleRec}
 \In rec=-2 (N+1) (N+2)^2 I[N]
 -(N+2) \big(
        -6 N^2-28 N-32\big) I[N+1]\newline
  \hspace*{2cm}+\big(-6 N^3-50 N^2-136 N-120\big) I[N+2]
-(-N-2) (N+4) (2 N+8) I[N+3]
==-\frac{4 (N+2)}{3 (N+3)};\\
\end{mma}
\noindent (loaded into Mathematica) and the initial values $I(1)=5$, $I(2)=\frac{130}{27}$, $I(3)=\frac{169}{36}$. Loading in the summation package \text{Sigma} into Mathematica, one can solve the recurrence in terms of nested hypergeometric sum expressions as follows.

\begin{mma}
\In << Sigma.m \\
\Print \LoadP{Sigma - A summation package by Carsten Schneider
\copyright\ RISC-Linz}\\
\notag
\MLabel{MMA:SolveRec}
\In recSol=SolveRecurrence[rec,I_{1,-3}[N],IndefiniteSummation\to False]\\
\vspace*{-0.2cm}
\Out     \{\{0 , \frac{1}{-N-1}\},
         \{0 , -\frac{\sum_{i=1}^N 1}{N+1}\},
         \{0 , -\frac{\ds
        \sum_{i=1}^N 
                \sum_{j=1}^{i} \frac{1}{j \big(
                        1+j\big)}}{N+1}\},
         \{1 , -\frac{\ds
        \sum_{i=1}^N 
                \sum_{j=1}^{i} \frac{\ds
                \sum_{k=1}^{j} \frac{2}{3}}{j \big(
                        1+j\big)}\big)}{N+1}\}\}\\
\end{mma}
\noindent Here the first three entries are linearly independent solutions of the homogeneous version of the recurrence and the last entry is a particular solution of the recurrence itself.\\
Within \texttt{Sigma} a strong toolbox has been developed to simplify these solutions by flattening the sums optimally and by finding denominators with minimal degrees within the setting of difference rings~\cite{Summation,CASummation}. 
In particular, the simplified expressions are built by sums that are algebraically independent~\cite{Schneider:10b}.
These features can be activated by dropping \texttt{IndefiniteSummation$\to$False} in~\myIn{\ref{MMA:SolveRec}}.
For the class of harmonic sums, generalized harmonic sums, cyclotomic harmonic sums and binomial nested sums these features are also available within the package  \begin{mma}
\In << HarmonicSums.m \\
\Print \LoadP{HarmonicSums by Jakob Ablinger
\copyright\ RISC-Linz}\\
\notag
\end{mma}
\noindent For such sums the rather involved difference ring theory can be avoided, and one can calculate very efficiently the simplified representation as follows:
\begin{mma}\MLabel{MMA:ReduceToBasis}
 \In recSol=ReduceToBasic[TransformToSSums[recSol],Dynamic\to Automatic]\\
 \Out     \{\{0 , \frac{1}{-N-1}\},
         \{0 , -\frac{N}{N+1}\},
         \{0 , \frac{1}{(N+1)^2}
        +\frac{\ds
        S_1(N)}{N+1}\},
         \{1 , \frac{2 \big(
                N^2+N-1\big)}{3 (N+1)^2}
        -\frac{\ds 2 (N+2)}{3 (N+1)}S_1(N)\}\\
 \end{mma}
 
\smallskip 
 
\noindent\textit{Remark.}  Internally, only the so-called basis-sums remain that cannot be eliminated by relations induced by the underlying quasi-shuffle algebra. For harmonic sums these ideas are worked out in~\cite{Blumlein:2003gb} and have been extended for cyclotomic sums, generalized harmonic sums and nested binomial sums~\cite{HarmonicSums,Ablinger:2013cf,Ablinger:2011te,Ablinger:2014bra}. We remark further that the basis sums produce sequences which are algebraically independent~\cite{AIHarmonicSums}.
 
\smallskip
 
Summarizing, the solution set of~\myOut{\ref{MMA:ReduceToBasis}} is given by the set
$$\{c_1\,\tfrac{-1}{N+1}+c_2\tfrac{-N}{N+1}+c_3\Big(\tfrac{1}{(N+1)^2}
        +\tfrac{
        S_1(N)}{N+1}\Big)+\tfrac{2(
                N^2+N-1)}{3 (N+1)^2}
        -\tfrac{2 (N+2)}{3 (N+1)}S_1(N)| c_1,c_2,c_3\in\KK\}$$
of nested hypergeometric sum expressions. The initial values $I(1),I(2),I(3)$ can be fulfilled with $c_1=-\frac{49}{9}, c_2=-\frac{41}{9}, c_3=-\frac{2}{3}$ which yields
\begin{equation}\label{Equ:RecSolverSol}
I(N)=\frac{59 N^2+120 N+49}{9 (N+1)^2}
-\frac{2 (N+3) S_1({N})}{3 (N+1)}.
\end{equation}
Hence we have shown that $I(N)$ can be calculated for $N\geq0$ by a nested hypergeometric expression.

\subsection{Finding Laurent series solutions of linear difference equations}

During the calculation of a Feynman integral $I(N)$ one often obtains linear recurrences for 
$I(N)$ 
depending on a dimensional parameter $\ep$ with $D=4+\ep\in\RR$. In lucky situations one finds a nested hypergeometric sum representation of $I(N)$ using the recurrence solver of Subsection~\ref{Subsec:RecSolver} with $\ep\in\KK$. However, in most cases one fails to find any solution of the given recurrence in terms of nested hypergeometric sum expressions, but one finds an $\ep$-expansion whose coefficients can be represented by nested hypergeometric expressions using the following algorithmic machinery~\cite{Blumlein:2010zv}.

\medskip

\noindent\textbf{Theorem~2.}
Suppose that the sequence $\langle I(N)\rangle_{N\geq0}\in\KK((\ep))^{\NN}$ with~\eqref{Equ:IExpansion}
is a solution of the difference equation 
\begin{equation}\label{Equ:RecExpansion}
a_0(\ep,N)I(N)+a_1(\ep,N)I(N+1)+\dots+a_d(\ep,N)I(N+d)=r(N)
\end{equation}
for explicitly given $a_0(\ep,N),\dots,a_d(\ep,N)\in\KK(\ep,N)$ and for a sequence $\langle r(N)\rangle_{N\geq0}\in\KK((\ep))^{\NN}$ with 
\begin{equation}\label{Equ:RecRExpansionOrg}
r(x)=\ep^or_{o}(N)+\ep^{o+1}r_{o+1}(N)+\ep^{o+2}r_{o+2}(N)+\dots
\end{equation}
Here we assume\footnote{\label{ftn:aiProp}This assumption can be always guaranteed by multiplying an appropriate factor $\ep^{s}$ with $s\in\ZZ$ on both sides of~\eqref{Equ:IExpansion}.} that the $a_i(\ep,N)|_{\ep\to0}$ do not introduce poles for $0\leq i\leq d$ and that not all $a_i(0,N)$ are zero. Then for any $u\in\ZZ$ one can determine an $m\in\NN$ with the following property.\\
If one is given the values $I_{j}(N)$ for all $o\leq j\leq u$ and $0\leq N\leq m$ and one is given for all $o\leq j\leq u$ nested hypergeometric sum expressions that calculate the values $r_{j}(N)$ for all $N\geq0$ (or at least from a certain point on), then one can decide algorithmically if for all $o\leq j\leq u$ there exist nested hypergeometric sum expressions that calculate the values $I_{j}(N)$ for all $N\geq0$ 
(or at least from a certain point on).

\medskip

\noindent\textit{Proof.} Let $u\in\ZZ$ and suppose that the $r_{o}(N),\dots,r_{u}(N)$ can be represented in terms of nested hypergeometric sum expressions.
We make the Ansatz~\eqref{Equ:IExpansion} with unknown coefficients $I_{j}(N)$ and plug them into~\eqref{Equ:RecExpansion}. Then the left and right hand sides are both Laurent series which are equal if and only if the coefficients agree. In particular, the lowest term must agree, i.e., we obtain the following constraint
\begin{equation}\label{Equ:ModRec}
a_0(0,N)I_{o}(N)+a_1(0,N)I_{o}(N+1)+\dots+a_d(0,N)I_{o}(N+d)=r_{o}(N)
\end{equation}
which is a linear recurrence of order
$d'=\max\{0\leq i\leq d| a_i(0,N)\neq0\}\geq0$.
Activating Theorem~1 to this recurrence with appropriately chosen initial values (given by Theorem~1) one can decide algorithmically if $I_o(N)$ is expressible by a nested hypergeometric sum expression. If such a representation is not possible, the theorem is proven. Otherwise, we take this representation and make the Ansatz~\eqref{Equ:IExpansion} with the known coefficient $I_{o}(N)$ and the unknown coefficients $I_{j}(N)$ ($j>o$) and plug them into~\eqref{Equ:RecExpansion}. Then by construction the coefficients  of $\ep^{o}$ on the left and right hand sides agree and the term of $\ep^o$ can be eliminated by subtracting it on both sides. Now one repeats this process for the next lowest term. In this way one can decide algorithmically if all $I_o(N),\dots,I_u(N)$ can be represented by nested hypergeometric sum expressions. In the process of this construction we choose $m\in\ZZ$ such that the used initial values are covered by $I_j(0),\dots,I_j(m)$ with $o\leq j\leq u$.

\medskip

\noindent\textit{Example.}
Take $\langle I(N)\rangle_{N\geq0}$ with the $\ep$-expansion~\eqref{Equ:IExpansion} of order $o=-3$ where the first two coefficients are determined by the initial values 
\begin{equation}\label{Equ:InitialI}
\IIntegral(1) = \tfrac{5}{\ep^3} -\tfrac{163}{12 \ep^2}+O(\ep^{-1}),\quad
  \IIntegral(2) =  \tfrac{130}{27 \ep^3} -\tfrac{695}{54 \ep^2} +O(\ep^{-1}),\quad
   \IIntegral(3) =  \tfrac{169}{36 \ep^3} -\tfrac{395}{32 \ep^2} +O(\ep^{-1})
\end{equation} 
and the linear recurrence 
\begin{mma}\MLabel{MMA:epRec}
\In recEp=-2 (N+1) (N+2) (2
+\ep
+N
) \IIntegral[N]-(N+2) \big(
        -32
        -7 \ep
        +2 \ep^2
        -28 N
        -5 \ep N
        -6 N^2
\big)\IIntegral[N+1]-\big(120
        +3 \ep
        -14 \ep^2
        -\ep^3
        +136 N
        +13 \ep N
        -4 \ep^2 N
        +50 N^2
        +4 \ep N^2
        +6 N^3
\big) \IIntegral[N+2]
+(2
-\ep
+N
) (4
+\ep
+N
) (8
+\ep
+2 N
) \IIntegral[N+3]\newline
\quad\quad==
\frac{1}{\ep^3}\frac{-4 (N+2)}{3 (N+3)}
 +\frac{1}{\ep^2}\Big[-\frac{2 (2 N+7) S_1}{3 (N+3)}
-\frac{2\big(
        4 N^4+35 N^3+101 N^2+105 N+25\big)}{3(N+1) (N+2) (N+3)^2}\Bigr]+O(\ep^{-1}).\\
\end{mma}
\noindent We seek to calculate a nested hypergeometric sum representation for $I_{-3}$, and $I_{-2}$, i.e., we set $u=-2$. Note that the expansion on the right hand side of~\myIn{\ref{MMA:epRec}} is sufficiently high expanded. The recurrence~\eqref{Equ:ModRec} in our concrete instance is precisely~\myIn{\ref{MMA:SimpleRec}} with $I_{-3}(N)=I(N)$. In addition, the initial values agree with~\eqref{Equ:InitialI}. Hence the found nested hypergeometric sum expression~\eqref{Equ:RecSolverSol} represents $I_{-3}(N)$. Continuing this process we can calculate the coefficient $I_{-2}(N)$. 
Within \texttt{Sigma} this calculation can be carried out automatically with the function call

\begin{mma}\MLabel{MMA:GenerateExpansion}
\In GenerateExpansion[recEp[[1]],\{Coefficient[recEp[[2]],\ep^{-3}],Coefficient[recEp[[2]],\ep^{-2}]\}, I[N],\{\ep,-3,-2\},\newline
\hspace*{6cm}\{\{ 5 , \frac{130}{27},\frac{169}{36}\},
 \{-\frac{163}{12},
   -\frac{695}{54},-\frac{395}{32}\}\},MinInitialValue\to1]\\
\Out \{\frac{59 N^2+120 N+49}{9 (N+1)^2}
-\frac{2 (N+3) S_1({N})}{3 (N+1)},-\frac{2(20 N^3+58 N^2+57 N+22)}{3(N+1)^3}
        +\frac{2 (N+2) (2 N-1) S_1}{3 (N+1)^2}
        -\frac{S_1^2}{N+1}
        -\frac{S_2}{N+1}\}\\
\end{mma}

\smallskip

\noindent\textit{Remark.} For further details and speed-ups we refer to~\cite{Blumlein:2010zv}.

\subsection{Finding Laurent series solutions of coupled systems of linear difference equations}\label{SubSec:CoupledRec}

We generalize Theorem~2 to solve coupled systems as follows~\cite{VLadders}.

\medskip 

\noindent\textbf{Theorem~3.}
Suppose that the sequences $\langle I_1(N)\rangle_{N\geq0}\rangle,\dots,\langle I_n(N)\rangle_{N\geq0}\rangle\in(\KK((\ep)))^{\NN}$ 
with
$$I_i(N)=\ep^{o}I_{i,o}(N)+\ep^{o+1}I_{i,o+1}(N)+\ep^{o+2}I_{i,o+2}(N)+\dots$$
are solutions of the coupled system of difference equations 
\begin{equation}\label{Equ:GenericCRS}
A_0 \left(\begin{matrix}I_1(N)\\ \vdots\\ I_n(N)\end{matrix}\right)
+A_1 \left(\begin{matrix}I_1(N+1)\\ \vdots\\ I_n(N+1)\end{matrix}\right)
\dots+A_d \left(\begin{matrix}I_1(N+d)\\ \vdots\\ I_n(N+d)\end{matrix}\right)=
\left(\begin{matrix}r_1(N)\\ \vdots\\ r_n(N)\end{matrix}\right)
\end{equation}
for explicitly given $n\times n$ matrices $A_0,\dots,A_d$ with entries from $\KK(\ep,N)$ and for some sequences $\langle r_1(N)\rangle_{N\geq0},\dots,\langle r_n(N)\rangle_{N\geq0}\in\KK((\ep))^{\NN}$ with
\begin{equation}\label{Equ:RiExpr}
r_i(N)=\ep^or_{o,i}(N)+\ep^{o+1}r_{o+1,i}(N)+\ep^{o+2}r_{o+2,i}(N)+\dots
\end{equation}
Then for any $u_1,\dots,u_n\in\ZZ$ one can determine $m_1,\dots,m_n\in\NN$ $v_1,\dots,v_n\in\ZZ$ and $w_1,\dots,w_n\in\ZZ$ with the following property.\\
If one is given the values $I_{i,j}(N)$ for all $1\leq i\leq n$, $o\leq j\leq v_i$ and $0\leq N\leq m_i$ and one is given for all $1\leq i\leq n$ and $o\leq j\leq w_i$  nested hypergeometric sum expressions that calculate the values $r_{i,j}(N)$ for all $N\geq0$ (or at least from a certain point on), then one can decide algorithmically if for all $1\leq i\leq n$ and $o\leq j\leq u_i$ there are nested hypergeometric sum expressions that calculate the values $I_{i,j}(N)$ for all $N\geq0$ 
(or at least from a certain point on).

\medskip

\noindent\textit{Proof.} The algorithmic steps can be summarized as follows. First, we transform the coupled system~\eqref{Equ:GenericCRS} to a first order system as explained in~\cite{VLadders}. Then we can apply any decoupling algorithm from~\cite{UNCOUPL} to uncouple the system, e.g., w.r.t.\ $I_1(N)$. In our implementation we chose Z\"urcher's algorithm~\cite{Zuercher:94} implemented in the package~\texttt{OreSys}~\cite{OreSys}.
In the generic case one obtains one linear recurrence  
in $I_1(N)$ which is of the form~\eqref{Equ:RecRExpansionOrg}. We can assume that the evaluations $a_i(0,N)$ are possible and that not all $a_i(0,N)$ are zero (see Footnote~\ref{ftn:aiProp}). In addition, the decoupling algorithm expresses the remaining sequences $I_2(N),\dots,I_{n}(N)$ by a linear combination of the shifted versions of $I_{1}(N)$ and shifted versions of the~\eqref{Equ:RiExpr}.\\
A subtle point is to which order $\ep^{w_i}$ the expressions in~\eqref{Equ:RiExpr} should be expanded. To extract this knowledge, we decouple the system by considering $r_1(N),\dots,r_n(N)$ first as unspecified sequences. Then analysing the corresponding output gives an upper bound for the $w_i$; details on these aspects can be found in~\cite{VLadders}. Since the description of the $I_2(N),\dots,I_n(N)$ is given in terms of a linear combination of the shifted versions of $I_1(N)$, it might be necessary to expand $I_1(N)$ higher than $u_1$. E.g., if $\ep^l I_1(N+i)$ with $l<0$ is one of the components. Analysing these combinations yields the required order $\nu_1$. Now we are ready to apply Theorem~2 to calculate the values $I_{1,i}(N)$ with $o\leq i\leq\nu_1$ by means of a linear recurrence. In this process we determine that the first $m_1$ initial values are needed\footnote{In all our examples $m_1$ agreed with the recurrence order plus some extra points induced by the physical problem.}. If one fails to get the representation of $I_1(N)$ in terms of nested hypergeometric sum expressions, the theorem is proven. Otherwise, by the properly chosen $\nu_1$, the  $w_1,\dots,w_n$ and $m_1$, this yields also a nested hypergeometric sum representation of the $I_2(N),\dots,I_{n}(N)$.\\ 
In the degenerated case, the decoupling algorithm provides several scalar linear recurrences, say in the $I_1(N),\dots,I_l(N)$, and the remaining $I_{l+1}(N),\dots,I_n(N)$ are expressed by a linear combination of the shifted versions of the $I_1(N),\dots,I_l(N)$ and the $r_i(N)$. Applying Theorem~2 with the corresponding $w_i,\nu_i,m_i$ (as described for the generic case) $l$~times leads to the desired result.

\smallskip

\noindent\textit{Example.}
Consider the sequences $I_1(N),I_2(N),I_3(N)$ which are solutions of the coupled system~\eqref{Equ:GenericCRS} with $d=1$ and $n=3$ where
\begin{align*}
A_0=\left(\begin{smallmatrix}
         N+1 & 0 & 0 \\
         \ep (3 \ep+2) & -2 (3 \ep+1) & -2 (-1
        +\ep
        -2 N
        ) \\
         -\ep (3 \ep+2) & 2 (3
        +3 \ep
        +2 N
        ) & 2 (\ep+1) \\
        \end{smallmatrix}\right),&&
A_1=\left(\begin{smallmatrix}
         -2
        -\ep
        -N
         & 2 & 0 \\
         -2 \ep (3 \ep+2) & 2 (5 \ep+2) & 4 (-1
        +\ep
        -N
        ) \\
         0 & -2 (4
        +\ep
        +2 N
        ) & 0 \\
        \end{smallmatrix}\right)
\end{align*}
and
\begin{equation}\label{Equ:riConcrete}
\begin{split}
r_1(N)&=(-\tfrac{4 (N+3)}{3 (N+2)}\ep^{-3}+\big(\tfrac{2}{3}\tfrac{6 N^3+29 N^2+45 N+21}{(N+1) (N+2)^2}
      -\tfrac{2 (2 N+3) S_ 1({N})}{3 (N+2)}\Big)\ep^{-2}+O(\ep^{-1}),\\
r_2(N)&=-\tfrac{8}{3}\ep^{-3} +\Big(\tfrac{4 (3 N+1)}{3 (N+1)}
        -\tfrac{8S_ 1({N})}{3}\Big)\ep^{-2}+O(\ep^{-1}),\\
r_3(N)&= \tfrac{8}{3}\ep^{-3} + \Big(\tfrac{-4 (3 N+1)}{3 (N+1)}
        +\tfrac{8S_ 1({N})}{3}\Big)\ep^{-2}+O(\ep^{-1})).
\end{split}
\end{equation}
In particular, we are given the initial values~\eqref{Equ:InitialI} for $I_1(N):=I(N)$ 
We want to derive the $\ep$-expansions for the $I_i(N)$ up to the order $u_i=-2$ for $1\leq i\leq 3$. Uncoupling the system (first with generic right hand sides) shows that $w_1=w_2=w_3=-2$, i.e., the $\ep$-expansions in~\eqref{Equ:riConcrete} are sufficiently high expanded. In particular, we obtain the linear recurrence~\myIn{\ref{MMA:epRec}} with $I(N)=I_1(N)$ and can express $I_2(N)$ and $I_3(N)$ by the shifted versions of $I_1(N)$ and $r_1(N),r_2(N),r_3(N)$. Summarizing, the coefficients $I_{1,-3}(N)$ and $I_{1,-2}$ are computed in \myOut{\ref{MMA:GenerateExpansion}}. This yields the needed information to calculate the $\ep$-expansions of $I_2(N)$ and $I_3(N)$ up to $\ep^{-2}$.

The full machinery can be summarized after loading in the packages
\begin{mma}
\In << OreSys.m \\
\Print\LoadP{OreSys by Stefan Gerhold (optimized by C. Schneider)
\copyright\ RISC-Linz}\\
\In << SolveCoupledSystem.m \\
\Print\LoadP{SolveCoupledSystem by Carsten Schneider
\copyright\ RISC-Linz}\\
\end{mma}
\noindent First, we execute the following command from the package \texttt{SolveCoupledSystem}: 
\begin{mma}\MLabel{MMA:AnalyzeCoupledRecSystem}
\In AnalyzeCoupledRecSystem[\{(A_0.\{I_1[N],I_2[N],I_3[N]\}+A_1.\{I_1[N],I_2[N],I_3[N]\},\{I_1[N],I_2[N],I_3[N]\},
\ep,{-2,-2,-2}]\\
\Out \{\{\{I_1[N],3,-2\}\},\{-2,-2,-2\},\{\}\}\\
\end{mma}
\noindent This means that one can solve the system by providing the three consecutive initial values of $I_1(N)$ up to order $\nu_1=-2$ (the starting point depends usually on the physical problem) and that one needs the $\ep$-expansions of $r_1(N),r_2(N),r_3(N)$ up to the orders $m_1=m_2=m_3=-2$. Providing the required information, we execute
\begin{mma}
\In coupledSys=A_0.\{I_1[N],I_2[N],I_3[N]\}+A_1.\{I_1[N],I_2[N],I_3[N]\}-\{r_1[n],r_2[n],r_3[N]\};\\
\In\label{MMA:SolveCoupledRecSystem} SolveCoupledRecSystem[coupledSys,\newline 
\{I_1[N],I_2[N],I_3[N]\}, \ep,-3,\{-2,-2,-2\},\{I_1[N],1,\{\tfrac{5}{\ep^3} -\tfrac{163}{12 \ep^2},\tfrac{130}{27 \ep^3} -\tfrac{695}{54 \ep^2},\tfrac{169}{36 \ep^3} -\tfrac{395}{32 \ep^2}\}\}]\\
\Out \Big\{ \frac{1}{\ep^3}\Big(\frac{4 \big(
                3 N^2+6 N+4\big)}{3 (N+1)^2}
        +\frac{4 S_1}{3 (N+1)}
        \Big)
        +\frac{1}{\ep^2}\big(-\frac{2(20 N^3+58 N^2+57 N+22)}{3(N+1)^3}
        +\frac{2 (N+2) (2 N-1) S_1}{3 (N+1)^2}
        -\frac{S_1^2}{N+1}
        -\frac{S_2}{N+1}
        \big),\newline
  \hspace*{0.4cm}\frac{4}{3 \ep^3} -\frac{2}{\ep^2},   
         \frac{8}{3 \ep^3} +\frac1{\ep^2}\big(-\frac{4 \big(
                4 N^2+7 N+2\big)}{3 (N+1)^2}
        +\frac{4 (N+2) S_1}{3 (N+1)}
       \big)\Big\}\\
\end{mma}
\noindent and obtain the $\ep$-expansions of $I_1(N),I_2(N)$ and $I_3(N)$ up to the orders $-2,-2$ and $-2$, respectively.

\subsection{Finding power series solutions of coupled systems of linear differential equations}

Finally, we are ready to present our differential equation solver for coupled systems.

\medskip

\noindent\textbf{Theorem~4.}
Suppose that the power series $\hat{I}_1,\dots,\hat{I}_n\in\KK((\ep))[[x]]$ with
\begin{equation}\label{IHatExp}
\hat{I}_i(x)=\sum_{N=0}^{\infty}\Big(I_{i,o}(N)\,\ep^{o}+I_{i,o+1}(N)\,\ep^{o+1}+I_{i,o+2}(N)\,\ep^{o+2}+I_{i,o+3}(N)\,\ep^{o+3}+\dots\Big)x^N
\end{equation}
for some common $o\in\ZZ$
are solutions of the coupled system of differential equations\footnote{Here $D_x$ stands for the derivative operator.} 
\begin{equation}\label{Equ:GenericCDS}
A_0 \left(\begin{matrix}\hat{I}_1(x)\\ \vdots\\ \hat{I}_n(x)\end{matrix}\right)
+A_1 D_x\left(\begin{matrix}\hat{I}_1(x)\\ \vdots\\ \hat{I}_n(x)\end{matrix}\right)
\dots+A_{\delta} D_x^{\delta}\left(\begin{matrix}\hat{I}_1(x)\\ \vdots\\ \hat{I}_n(x)\end{matrix}\right)=
\left(\begin{matrix}\hat{r}_1(x)\\ \vdots\\ \hat{r}_n(x)\end{matrix}\right)
\end{equation}
for explicitly given $n\times n$ matrices $A_0,\dots,A_{\delta}$ with entries from $\KK(x)$ and for some $\hat{r}_1(x),\dots,\hat{r}_u(x)\in\KK((\ep))[[x]]$ with
\begin{equation}\label{Equ:RxDef}
\hat{r}_i(x)=\sum_{N=0}^{\infty}\Big(r_{i,o}(N)\,\ep^{o}+r_{i,o+1}(N)\,\ep^{o+1}+r_{i,o+2}(N)\,\ep^{o+2}+r_{i,o+3}(N)\,\ep^{o+3}+\dots\Big)x^N.
\end{equation}
Then for any $u_1,\dots,u_n\in\ZZ$ one can determine $m_1,\dots,m_n\in\NN$ and $w_1,\dots,w_n\in\ZZ$ with the property as stated in Theorem 3.

\medskip

\noindent\textit{Proof.} This result follows straightforwardly by holonomic closure properties.
Namely, take the $n$ partial linear differential equations in the $\hat{I}(x)$ and clear denominators by multiplying them with an appropriate polynomial from $\KK[\ep,x]$. Take one of the terms of the $n$ equations, say $a(\ep,x)\,D^k \hat{I}_i(x)$ with $0\leq k\leq\delta$, $1\leq i\leq n$ and $a(\ep,x)\in\KK[\ep,x]$. With the Ansatz~\eqref{IHatExp} we get
\begin{align*}
a(\ep,x)\,D^k \hat{I}_i(x)=&a(\ep,x) D^k\sum_{N=0}^{\infty}\Big(I_{i,o}(N)\,\ep^{o}+I_{i,o+1}(N)\,\ep^{o+1}+I_{i,o+2}(N)\,\ep^{o+2}+\dots\Big)x^N\\
=&a(\ep,x) \sum_{N=0}^{\infty}\Big(I_{i,o}(N)\,\ep^{o}+I_{i,o+1}(N)\,\ep^{o+1}+I_{i,o+2}(N)\,\ep^{o+2}+\dots\Big)\prod_{j=0}^{k-1}(N-j)x^{N-k}.
\end{align*}
Now we plug all these terms and~\eqref{Equ:RxDef} into the $n$ equations. Doing coefficient comparison w.r.t.\ $x^N$ leads to a coupled system of linear difference equations in the $I_1(N),\dots,I_n(N)$ and the $r_1(N),\dots,r_n(N)$. Performing an appropriate shift in $N$ yields a coupled system of the form~\eqref{Equ:GenericCRS} with the right hand sides $r'_i(N)$ which depend linearly on the $r_i(N)$ and their shifted versions from~\eqref{Equ:RxDef}. Note that the coefficients in $r'_i(N)$ can be expressed in terms of nested hypergeometric sum expressions since the coefficients in $r_i(N)$  can be expressed in terms of nested hypergeometric sum expressions. Thus we can apply Theorem~3 and obtain the claimed result.

\medskip

\noindent\textit{Example.}
Consider the integrals $\hat{\IIntegral}_1(x),\hat{\IIntegral}_2(x),\hat{\IIntegral}_3(x)$ given by
$\hat{I}_1(x) = J_6(1,1;x)$, $\hat{I}_2(x) = J_6(2,1;x)$ and $\hat{I}_3(x) = J_6(1,2;x)$
where
\begin{equation}\label{Equ:J6}
J_6(\nu_2,\nu_4;x) =
\int \frac{d^Dk_1}{(2 \pi)^D}  \frac{d^Dk_2}{(2 \pi)^D} \frac{d^Dk_3}{(2 \pi)^D} \,\,
\frac{1}{P_2^{\nu_2} P_4^{\nu_4} P_5 P_7 P_8 P_{10}}
\end{equation}
with the propagators
$P_2=(k_1-p)^2-m^2$, 
$P_4=(k_2-p)^2-m^2$, 
$P_5=k_3^2-m^2$, 
$P_7=(k_3-k_2)^2$, 
$P_8=(k_1-k_2)^2$ and
$P_{10}=1-x \Delta.k_1$.
Note that $\hat{I}_i(x)$ can be given in the power series representation~\eqref{IHatExp} and the main task is to determine the coefficients $I_{i,j}(N)$. 
The IBP algorithm of \texttt{Reduze~2}~\cite{Reduze2} delivers e.g. the coupled 
system~\eqref{Equ:GenericCDS} 
with $n=3$ and $\delta=1$ where $A_1$ is the $3\times3$ identity matrix and
$$A_0=-\left(
        \begin{matrix}
         \frac{\ds 1
        +\ep
        -x
        }{\ds (x-1) x}  & \frac{\ds -2}{\ds (x-1) x} & 0\\
         \frac{\ds -\ep (3 \ep+2) (x-2)}{\ds4 (x-1) x}  & \frac{\ds -2
        -5 \ep
        +x
        +3 \ep x
        }{\ds 2(x-1) x}& \frac{\ds (-2 \ep
        -x
        +\ep x
        )}{\ds 2(x-1) x}  \\
         \frac{\ds \ep (3 \ep+2)}{\ds 4 (x-1)}  & \frac{\ds 2
        +\ep
        -3 x
        -3 \ep x
        }{\ds 2(x-1) x} & -\frac{\ds \ep+1}{\ds 2 (x-1)}
        \end{matrix}\right).$$
Furthermore, the $r_i(x)$ are given by a linear combination of master integrals where each one has a power series representation of the form~\eqref{IHatExp}. In particular, one can use symbolic summation tools~\cite{Summation,CASummation} to calculate the coefficients of the coefficients (up to a certain modest order in $\ep$) in terms of nested hypergeometric sum expressions. 
For other situations multiple integration methods \cite{Integration} are the appropriate tool. Thus the $\hat{r}_i(x)$ have a power series representation of the form~\eqref{Equ:RxDef} where the $r_{i,j}(N)$ can be given (up to a certain modest $\ep$-order) explicitly in terms of nested hypergeometric sum expressions.\\ 
Now we activate our machinery. By holonomic closure properties we obtain precisely the coupled difference system from Example~\ref{SubSec:CoupledRec}. Taking the data from~\myOut{\ref{MMA:AnalyzeCoupledRecSystem}} we calculate the required $r_{i,j}(N)$ by means of symbolic summation. Furthermore, we calculate the initial values~\eqref{Equ:InitialI} by exploiting the $\alpha$-parametrization of the integrals; for further details on this method we refer 
to~\cite{Ablinger:2014uka}.
Note that in other situations we also used our summation tools, provided a reasonable sum representation has been derived. Finally, we activate the function call~\myIn{\ref{MMA:SolveCoupledRecSystem}} to get the final result. 

\smallskip

\noindent\textit{Remark.} The integrals $\hat{I}_1(x) = J_6(1,1;x)$, $\hat{I}_2(x) = J_6(2,1;x)$ and $\hat{I}_3(x) = J_6(1,2;x)$ with~\eqref{Equ:J6} themselves are master integrals produced by \texttt{Reduze 2}~\cite{Reduze2} in order to calculate the diagram\footnote{The graph has been produced by {\tt Axodraw}~\cite{Vermaseren:1994je}.}
\begin{equation}\label{Equ:D12}
\vcenter{\hbox{\includegraphics[width=4cm]{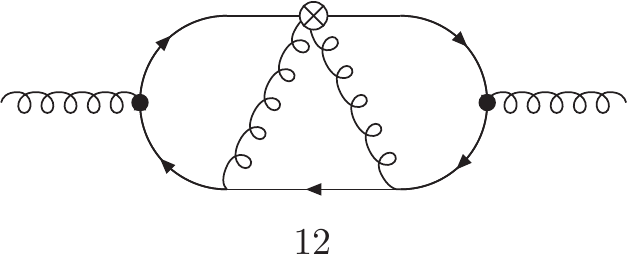}}}
\end{equation}
in~\cite{VLadders}. There we calculated the $\ep$-expansion up to order $3$ (and not just to order $-2$) in terms of 40 harmonic sums up to weight 7. The total calculation time was 229 seconds. The most complicated coupled system for diagram~\eqref{Equ:D12} had dimension $n=4$. Interestingly enough, the right hand sides of~\eqref{Equ:GenericCRS} can be given in terms of generalized harmonic sums only, but the solution is given in terms of nested binomial sums. This strongly indicates that straightforward tactics, like transforming the system to a particular shape and reading off the solutions, are not sufficient for such systems.\\
Note that the obtained results in $N$-space presented in~\cite{VLadders} and also in~\cite{DiffCalculations,Ablinger:2014uka} can be transformed to $x$-space by using iterated integral representations over general alphabets, generalizing in parts Poincar\'{e} iterated integrals~\cite{Ablinger:2014bra}.

\section{Conclusion}

We worked out a solver for coupled systems of linear differential equations where in one stroke
\begin{enumerate}
 \item the coefficients of the formal power series solution are expanded in its $\ep$-expansion;
 \item the coefficients of the $\ep$-expansion are given up to the desired order in terms nested hypergeometric expressions.
\end{enumerate}
More precisely, we presented a complete algorithm which either provides such a solution or proves that such a representation is not possible. These algorithms have been implemented in the new Mathematica package \texttt{SolveCoupledSystem} which is based on the packages \texttt{Sigma} and \texttt{HarmonicSums}.

In concrete calculations IBP-techniques provide a recursively defined system of coupled equations. In~\cite{VLadders} a new method has been worked out in order to treat these recursive systems fully automatically by using the algorithms presented in this article.

\vspace*{3mm}\noindent
{\bf Acknowledgement.}
We would like to thank A.~Behring and A.~von Manteuffel for discussions. 


\end{document}